\documentclass[aps,prd,eqsecnum,floats,floatfix,twocolumn,nofootinbib,
showpacs]{revtex4}

\usepackage{graphicx}
\usepackage{bm}
\usepackage{amssymb}

\newcommand{\eqref}[1]{(\ref{#1})}
\newcommand{\beq}{\begin{equation}}
\newcommand{\eeq}{\end{equation}}
\newcommand{\beqa}{\begin{eqnarray}}
\newcommand{\eeqa}{\end{eqnarray}}

\newcommand{\Lieshift}{{\cal L}_{\vec N}}

\begin{document}

\title{Constraint Damping in First-Order Evolution Systems for Numerical 
Relativity}

\author{Robert Owen}

\affiliation{Theoretical Astrophysics 130-33, California Institute of
Technology, Pasadena, CA 91125}

\date{\today}

\begin{abstract}
A new constraint suppressing formulation of the Einstein evolution
equations is presented, generalizing the five-parameter
first-order system due to Kidder, Scheel and Teukolsky (KST).  
The auxiliary fields, introduced to make the KST system first-order, are given 
modified evolution equations designed to drive constraint violations toward 
zero.  The algebraic structure of the new system is
investigated, showing that the modifications preserve the hyperbolicity of the 
fundamental and constraint evolution equations.  The
evolution of the constraints for pertubations of flat spacetime is
completely analyzed, and all finite-wavelength constraint modes are 
shown to decay exponentially
when certain adjustable parameters satisfy appropriate 
inequalities.
Numerical simulations of a single Schwarzschild black hole are presented, 
demonstrating the effectiveness of the new constraint-damping modifications.

\end{abstract}


\maketitle

\section{Introduction}

Numerical relativity has recently undergone a revolution.  Multiple 
research groups, using a variety of mathematical and computational formalisms, 
have produced consistent pictures of the late inspiral and 
coalescence of binary black hole systems~\cite{Pretorius2005a, 
Campanelli2006a, Campanelli2006b, Campanelli2006c, Baker2006a, Baker2006b, 
Scheel2006, Bruegmann2006}, a goal that until 
recently seemed remote.  The community is opening a theoretical 
window on issues fundamental to gravitational wave astrophysics, but much work 
still remains to be done.

State of the art simulations have provided gravitational waveforms due to the 
last several orbits of a binary black hole system, and its eventual 
coalescence and ringdown.  This is an extraordinary achievement, but 
numerical relativity may need to handle well over a dozen orbits accurately 
before a seamless transition can be made from post-Newtonian analysis.  This 
requires an exceptionally stable evolution scheme, and also an extremely 
efficient one, if the simulations are to be done in a timely manner.

Numerical simulations can become unstable for a variety of reasons, some 
purely numerical (such as a poor choice of algorithm), some purely 
mathematical (such as ill-posedness of the continuum mathematical problem), 
and some a combination of the two.  The subject of this paper is 
an instability of this last type: the exponential growth of the constraint 
fields under free evolution -- an instability of the continuum evolution 
equations, seeded by numerical errors.

A variety of methods exist to deal with such instabilities.  One very well 
established method is known as constrained evolution~\cite{StPi85, AbEv92, 
ch93, AbEv93, ACST94, Choptuik2003, Choptuik2003a, Rinne2005}, in which some 
subset of 
the dynamical fields are integrated in time using the evolution equations, and 
others are obtained by solving the constraints after each time step.  This 
separation of the fields into an ``evolved'' family and a ``constrained'' 
family is normally guided by some sort of symmetry.  A related method, known 
as constraint projection~\cite{Anderson2003, Holst2004}, places all fields on 
an equal footing, freely integrating everything using the evolution equations, 
then periodically ``projecting'' the fields down to the constraint-satisfying 
subset of the solution space.  It appears that this method can be quite 
robust in practice, but it can also be technically demanding, requiring 
the repeated solution of nonlinear elliptic equations.

A preferable approach for dealing with these instabilities, whenever possible, 
is to remove them from the evolution system at the continuum level, before 
they reach the numerical code.  This type of effort is often referred to 
as ``constraint damping.''  
It is possible to change the evolution equations without changing the
physics they represent.  For instance, coordinates can be chosen freely on
the simulated spacetime, and indeed, careful choices of gauge have been
shown to have a strong effect on the stability of simulations,
particularly in the BSSN 
system~\cite{Alcubierre2002, Yo2002, Gundlach2006, vanMeter2006}.  
Another, perhaps more drastic method to stabilize constraints involves
extending the family of evolved fields.  It is possible, with some care, 
to introduce fields whose evolution will naturally lead to the 
presence of friction terms in the implied constraint evolution system.  
Systems of this form have come to be referred to as ``$\lambda$ systems'' 
in the relativity literature, after the pioneering work of Brodbeck 
et al.~\cite{Brodbeck1999, Siebel2001}.

The constraint damping of Ref.~\cite{Brodbeck1999} utilizes the freedom 
to substitute constraint equations into the evolution equations.  In the 
physical situation where the constraints are all satisfied, the equations 
are unchanged.  However in the numerical situation where the constraint 
satisfaction is at best approximate, these substitutions can have a profound 
effect on the structure of the evolution system and the stability of its 
constraints.  In Ref.~\cite{Kidder2001}, Kidder, Scheel and Teukolsky 
added terms, proportional to the constraints, to a first-order representation 
of the Arnowitt-Deser-Misner (ADM) evolution equations~\cite{ADM} (starting 
from the form advocated by York~\cite{york79}).  
With these modifications, they were able to change the principal part of the 
evolution system.  When the free parameters satisfy certain inequalities, 
the evolution system becomes strongly, or symmetric, 
hyperbolic~\cite{Kidder2001, Lindblom2002, Lindblom2003, Kidder2005}.  
The purpose of the present 
paper is to generalize the KST systems even further, to add more terms 
proportional to constraints into the evolution equations, but now with the 
goal of damping the constraints while preserving hyperbolicity.  It will be 
shown that this goal can largely be achieved, stabilizing all the constraints 
of the KST systems, without the need to introduce extra fields as in the 
``$\lambda$ system'' approach.

The possibility of constraint damping along these lines is now widely seen 
as a major advantage of the generalized harmonic formalism.
Pretorius~\cite{Pretorius2005c, Pretorius2005a}, building 
on Gundlach et al.~\cite{Gundlach2005}, introduced such a modification in his 
generalized harmonic evolution code, leading to the first 
ever simulation of a full orbit and merger of binary black holes.  
In Ref.~\cite{Lindblom2005}, this system was converted into an explicitly 
first-order, linearly degenerate, symmetric-hyperbolic form.  An extensive 
body of mathematical literature exists on systems of this form, and they are 
also very well suited to highly accurate multidomain pseudospectral 
collocation methods.  While this first-order generalized harmonic system is 
perfectly acceptable for numerical relativity, and indeed is now used for 
nearly all simulations currently being done by the 
Caltech and Cornell numerical 
relativity groups, there would be considerable value in implementing a KST 
system of comparable stability.  For one thing, the KST system involves just 
over half as many fields as the first-order generalized harmonic system, so 
it could provide a considerable improvement in code runtime.  The KST systems 
are also closer to the evolution systems used historically in numerical 
relativity.  Gauge is specified again in terms of (densitized) lapse and 
shift.  So a large body of research on gauge conditions can be more easily 
applied.  

As the standard KST systems already have adjustable 
parameters, it is an interesting question whether any of them 
can have constraint-damping properties.  Even without calculations, it is 
clear that the answer must be 
``no.''  With the generalized-harmonic system presented in 
Ref.~\cite{Lindblom2005}, 
two parameters are available, $\gamma_0$ and $\gamma_2$, which tune the 
stability of small constraint-violating perturbations of flat spacetime.  
The inverses of these parameters (up to factors of order 
unity) define timescales on which short-wavelength constraint modes will decay 
exponentially.  All of the free parameters of the standard KST systems are 
dimensionless, so they cannot fix any preferred timescale for the damping of 
perturbations of flat spacetime.  Of course in the full nonlinear phase space, 
the dimensions necessary for such a timescale can be provided by nontrivial 
features of a particular solution, for instance the mass of a black hole.  
Indeed, in Ref.~\cite{Scheel2002}, it was demonstrated that careful 
fine-tuning of the parameters can considerably extend the 
lifetime of a black hole simulation.  
This effect cannot really be considered constraint damping, though, as the 
optimum choice of parameters depends strongly on the details of the initial 
data, and in fact it must fail for small features in asymptotic regions, 
where the above flat space arguments begin to apply.

In the present paper, the KST systems will be generalized even further, 
introducing one new free parameter with dimensions of inverse-time, that can 
be considered a mechanism for constraint damping.  This 
generalization is directly analogous to methods used in the 
past~\cite{Holst2004,Lindblom2005} for controlling the stability of the 
constraints that appear when an evolution system that is second-order in 
spatial derivatives is reduced to first-order form by the introduction of 
auxiliary fields.  Such a constraint exists in the KST systems, and while 
our methods are only designed to control this particular constraint, 
the intricate coupling of the constraints, in their evolution, 
extends the constraint damping effect to all (finite wavelength) 
constraint modes, including the Hamiltonian and momentum constraints.

The format of this paper is as follows.  In Sec.~\ref{s:illustration} the 
intuitive mechanism behind the new constraint damping terms is sketched out 
in the context of a simple toy model.  In Sec.~\ref{s:ModKST}, analogous 
modifications are applied to the five-parameter KST evolution systems, 
and conditions on these modifications are noted to keep the resulting 
system hyperbolic.  In Sec.~\ref{s:const_hyp}, the hyperbolicity 
of the constraint evolution system is investigated.  In 
Sec.~\ref{s:ModeAnalysis} the effectiveness of these modifications 
on constraint-violating perturbations of flat spacetime is seen in 
detail.  In Sec.~\ref{s:params}, the available parameter freedom is 
summarized, and in Sec.~\ref{s:num_tests}, results are presented
of a few simple numerical simulations using this new evolution system.

\section{Illustration of a Simple Model System}
\label{s:illustration}

Before jumping into the full equations of general relativity, it would be 
instructive to outline the constraint damping idea in the context 
of the simplest hyperbolic system, the scalar 
wave equation:
\beq
\eta^{\mu \nu} \partial_\mu \partial_\nu \psi = 0,\label{e:dalembert}
\eeq
for a real scalar field $\psi$, where $\eta_{\mu \nu}$ is the Minkowski metric 
in Cartesian coordinates.  This equation involves only one field, and no 
constraints, but it is second-order in both space and time.  The second-order 
derivatives are removed by promoting the first derivatives of $\psi$ to 
independent 
fields.  The first time derivative (up to a conventional 
minus sign) will be denoted with the symbol $\pi$.  The wave equation now 
becomes the system:
\beqa
\partial_t \psi &=& - \pi, \label{e:psiev2O}\\
\partial_t \pi &=& - \delta^{ij} \partial_i \partial_j \psi, \label{e:piev2O}
\eeqa
where latin indices refer to the spatial coordinates of the chosen inertial 
frame.  
The one-form $\phi_i$, defined by a new constraint ${\cal C}_i$, can be used 
to remove second spatial derivatives from the system:  
\beqa
\partial_t \psi &=& - \pi,\label{e:psiev}\\
\partial_t \pi &=& - \delta^{ij} \partial_i \phi_j,\label{e:piev}\\
{\cal C}_i &:=& \partial_i \psi - \phi_i = 0\label{e:scalarconstraint}.
\eeqa
Before this can be useful for free evolution, an equation is needed 
for evolving $\phi_i$.  This equation is commonly derived by equating the 
spatial coordinate derivatives of both sides of Eq.~\eqref{e:psiev}, commuting 
partial derivatives on the left, and substituting the constraint to find:
\beq
\partial_t \phi_i = - \partial_i \pi\label{e:phievstd}.
\eeq
This equation closes the evolution system, and leaves us with a nice 
first-order symmetric-hyperbolic representation of the scalar wave equation.  
However the newly defined constraint field is only marginally stable.  As 
is easily verified by direct substitution of the constraint definition and the 
evolution equations, the constraint is now conserved:
\beq
\partial_t {\cal C}_i = 0.
\eeq
This shows that exact constraint satisfaction should be preserved within the 
domain of dependence of the initial data, but also that any violations that 
may arise will be preserved as well.

Because the constraint is linear in undifferentiated $\phi_i$, anything added 
to the right side of Eq.~\eqref{e:phievstd} will transfer directly to the 
evolution equation implied for the constraint.  For example, if the 
equation is changed to:
\beqa
\partial_t \phi_i &=& - \partial_i \pi + \gamma {\cal C}_i
\label{e:gamma_introduced}\\
&=& - \partial_i \pi + \gamma (\partial_i \psi - \phi_i), \label{e:phievmod}
\eeqa
for some constant $\gamma$, then the constraint-satisfying solution space is 
unchanged, but the evolution of the constraint becomes
\beq
\partial_t {\cal C}_i = - \gamma {\cal C}_i .
\eeq
Thus, with this method, the constraint can be damped (assuming hyperbolicity 
is preserved) exponentially on an arbitrary fixed timescale $\gamma^{-1}$.  As 
we will see when we discuss the KST system, the constraint damping effect can 
extend even farther than the constraint that appears in the reduction.  Even 
those constraints that exist before the reduction to first-order form can be 
damped.

The question of whether this modification preserves the clear 
symmetric hyperbolicity of the standard reduction is important, and a very 
simple argument shows that hyperbolicity is not affected.  If a linear change 
of variables is made (in other words, a change of basis on the vector bundle 
of dynamical fields), defining $\bar \pi := \pi - \gamma \psi$, then all 
modifications of the principal part of the fundamental evolution system 
disappear:
\beqa
\partial_t \psi &\simeq& 0,\\
\partial_t \bar \pi &\simeq& - \delta^{ij} \partial_i \phi_j,\\
\partial_t \phi_i &\simeq& -\partial_i \bar \pi
\eeqa
(the symbol ``$\simeq$'' means that all nonprincipal -- in this case, 
algebraic -- terms have been omitted).  This transformed system is exactly the 
same as the unmodified system at 
principal order, and is clearly symmetric-hyperbolic.  The existence of a 
positive-definite symmetrizing inner product is independent of 
the basis of dynamical fields.  
Indeed, the obvious symmetrizer for the transformed system,
\beqa
dS^2 &=& \Lambda d\psi^2 + d{\bar \pi}^2 + \delta^{ij} d\phi_i d\phi_j,\\
&=& \Lambda d\psi^2 + (d\pi - \gamma d\psi)^2 + \delta^{ij} d\phi_i d\phi_j,
\eeqa
when expressed in terms of $\pi$, is positive-definite (for positive 
$\Lambda$) and symmetrizes the untransformed system.  Therefore, this 
constraint-damped form of the scalar wave system is symmetric-hyperbolic for 
any fixed choice of the damping timescale.

\section{The Modified KST Evolution System}
\label{s:ModKST}

The Kidder-Scheel-Teukolsky~\cite{Kidder2001} evolution equations are 
a five-parameter\footnote{A twelve-parameter system also exists, 
employing redefinitions of the fundamental dynamical fields and associated 
constraint substitutions.  Here we will ignore this extra freedom.} 
generalization of the standard first-order representation of the classic 
ADM~\cite{ADM} equations, in the form advocated by York~\cite{york79}: 
\beqa
\partial_t g_{ij} - \Lieshift g_{ij} &=& - 2 N K_{ij} \label{e:evg}\\
\partial_t K_{ij} - \Lieshift K_{ij} &=& N R_{ij} + N (K K_{ij} 
- 2 {K_i}^k K_{kj}) \nonumber \\ & & - \nabla_i \nabla_j N . \label{e:evK}
\eeqa
The dynamical fields are $\{g_{ij}, K_{ij}\}$, the metric intrinsic to the 
slice of constant $t$, and the extrinsic curvature of its embedding in 
spacetime.  The gauge fields $\{N, N^i\}$, lapse and shift, determine the 
evolution of the coordinates.

The Ricci tensor $R_{ij}$ written above is that of the spatial metric 
$g_{ij}$, so it implicitly involves second spatial derivatives of $g_{ij}$.  
The evolution system can be reduced to first-order form by promoting the 
partial 
derivatives of the spatial metric functions to an independent (nontensorial) 
three-index field:
\beq
D_{kij} := \frac{1}{2} \partial_k g_{ij}.
\eeq
As long as $D_{kij}$, under its own evolution, properly represents 
$\partial_k g_{ij}/2$, it can be substituted for any derivatives of 
$g_{ij}$.  This renders the ADM system first-order.

This evolution system, like that for the scalar field, describes physics only 
when certain constraint fields vanish:
\beqa
{\cal C} &:=& \frac{1}{2} (R - K_{ij} K^{ij} + K^2), \label{e:HamC}\\
{\cal C}_i &:=& \nabla^j (K_{ij} - K g_{ij}), \label{e:MomC}\\
{\cal C}_{kij} &:=& \partial_k g_{ij} - 2 D_{kij}. \label{e:C3}
\eeqa
The {\em Hamiltonian} and {\em momentum} constraints, ${\cal C}$ and 
${\cal C}_i$, must vanish (in vacuum) throughout each spatial slice, according 
to the four Einstein equations not represented in Eq.~\eqref{e:evK}.  The 
three-index constraint ${\cal C}_{kij}$ vanishes when $D_{kij}$ properly 
represents $\partial_k g_{ij}/2$, in analogy with the constraint of 
the scalar field system.  

The new field, $D_{kij}$, is to be considered independent in 
free evolution.  An evolution equation must be defined for this field, 
one that is consistent with the satisfaction of the three-index constraint 
above.  The 
equation analogous to Eq.~\eqref{e:phievstd} is
\beqa
\hat \partial_0 D_{kij} &=& \frac{1}{2} \partial_k (\hat \partial_0 g_{ij}),\\
&=& - \partial_k (N K_{ij}) ,\label{e:evD}
\eeqa
where the shorthand $\hat \partial_0$ refers to the derivative, 
$\partial_t - {\cal L}_{\vec N}$, along the normal to the spatial 
slice\footnote{Note that $\hat \partial_0$, involving a Lie derivative, 
commutes with the partial derivative $\partial_k$.  It is this commutation 
that defines the action of the Lie derivative on the nontensorial field 
$D_{kij}$.}.

The evolution system has now been written in first-order form, and we can 
begin to ask 
about its hyperbolicity.  Kidder, Scheel and Teukolsky~\cite{Kidder2001} have 
shown that the above system can be rendered strongly hyperbolic with a few 
simple modifications. The first of these is commonly referred to as 
{\em densitization of the lapse}.  Rather than fixing $N$ directly, we fix a 
related field $Q$ defined by
\beq
N = g^{\gamma_0} \exp(Q),
\eeq
where $g$ is the determinant of $g_{ij}$ and $\gamma_0$ is a constant nonzero 
parameter.  The occurences of $\partial_k g_{ij}$ that then arise arise from 
the $\nabla_i \nabla_j N$ term in Eq.~\eqref{e:evK} are then replaced by 
$2 D_{kij}$.  The second modification required for hyperbolicity is the 
addition of terms to the evolution equations for $K_{ij}$ and $D_{kij}$, that 
are proportional to the constraints:
\beqa
\hat \partial_0 K_{ij} &=& ... + \gamma_1 N g_{ij} {\cal C} + 
\gamma_2 N g^{ab} {\cal C}_{a(ij)b},\label{e:kstK}\\
\hat \partial_0 D_{kij} &=& ... + \frac{1}{2} \gamma_3 N g_{k(i} 
{\cal C}_{j)} + \frac{1}{2} \gamma_4 N g_{ij} {\cal C}_k,\label{e:kstD}
\eeqa
where the ellipses refer to the right sides of Eqs.~\eqref{e:evK} 
and~\eqref{e:evD}.  The four-index object 
${\cal C}_{klij} := 2 \partial_{[k} D_{l]ij}$ used here can be thought of as 
another constraint, but it vanishes automatically whenever the three-index 
constraint vanishes.  The parameters $\{\gamma_0, ..., \gamma_4\}$ do not 
affect the physical solution space of the equations in any way, but they 
directly affect the principal part of the evolution system.  
In Ref.~\cite{Kidder2001}, Kidder, Scheel and Teukolsky determined sufficient 
conditions for 
these parameters that render the evolution system strongly hyperbolic.  
In Ref.~\cite{Lindblom2003} and an appendix of Ref.~\cite{Kidder2005}, 
these arguments 
were extended, and it was also made clear on what subset of the parameter 
space the equations satisfied the stronger condition of symmetric 
hyperbolicity.  

The focus of this paper is a further modification of Eq.~\eqref{e:kstD} 
along the same lines as that described in Sec.~\ref{s:illustration}.  Here 
the goal is to modify the evolution of the three-index constraint, which in 
the ordinary KST system is implied to evolve as
\beq
\hat \partial_0 {\cal C}_{kij} = - \gamma_3 N g_{k(i} {\cal C}_{j)} 
- \gamma_4 N g_{ij} {\cal C}_k.
\eeq
Note that the need for damping here is more dire than in the scalar field 
case.  Hyperbolicity requires $\gamma_3$ and $\gamma_4$ to be nonzero, so any 
violation of the momentum constraint will feed directly into the three-index 
constraint.  This is countered as in the previous section by including terms 
proportional to the three-index 
constraint in the evolution equation for $D_{kij}$.  As we will see in a 
moment, multiples of the traces, denoted 
${\cal C}^1_k := g^{ij} {\cal C}_{kij}$ and 
${\cal C}^2_j := g^{ki} {\cal C}_{kij}$, must be added in separately, so the 
resulting evolution equation is:
\beqa
\hat \partial_0 D_{kij} &=& ... + \frac{1}{2} \gamma_3 N g_{k(i} 
{\cal C}_{j)} + \frac{1}{2} \gamma_4 N g_{ij} {\cal C}_k \nonumber \\
&&+ \frac{1}{2} N \gamma_5 {\cal C}_{kij} + \frac{1}{2} N \gamma_6 
{\cal C}^1_k g_{ij} \nonumber \\
&&+ \frac{1}{2} N \gamma_7 {\cal C}^2_{(i}g_{j)k} + \frac{1}{2} N \gamma_8 
{\cal C}^1_{(i}g_{j)k} \nonumber \\
&&+ \frac{1}{2} N \gamma_9 {\cal C}^2_k g_{ij}.
\eeqa
The term proportional to 
$\gamma_5$ is analogous to the term proportional to $\gamma$ in 
Eq.~\eqref{e:gamma_introduced}.  The terms proportional to $\gamma_6$, 
$\gamma_7$, $\gamma_8$ and $\gamma_9$ are necessitated by the hyperbolicity 
conditions, which we now consider.

The principal part of this system is:
\beqa
{\hat \partial}_0 g_{ij} &\simeq& 0,\label{e:ppg}\\
{\hat \partial}_0 K_{ij} &\simeq& N[g^{ab} \delta^c_i \delta^d_j 
- (1+\gamma_2) g^{ad}\delta^b_{(i}\delta^c_{j)} \nonumber \\ 
& & - (1-\gamma_2) g^{bc}\delta^a_{(i}\delta^d_{j)} 
+ (1+2{\gamma_0}) g^{cd}\delta^a_{(i}\delta^b_{j)} \nonumber \\ 
& & - \gamma_1 g^{ad} g^{bc} g_{ij} + \gamma_1 g^{ab} g^{cd} g_{ij} ] 
\partial_a D_{bcd} ,\label{e:ppK}\\
{\hat \partial}_0 D_{kij} &\simeq& - N[ \delta^c_k \delta^a_i \delta^b_j 
- \frac{1}{2} \gamma_3 g^{ca} g_{k(i} \delta^b_{j)} \nonumber \\ 
& & - \frac{1}{2} \gamma_4 g^{ca} g_{ij} \delta^b_k 
+ \frac{1}{2} \gamma_3 g^{ab} g_{k(i} \delta^c_{j)} \nonumber \\ 
& & + \frac{1}{2} \gamma_4 g^{ab} g_{ij} \delta^c_k ] \partial_c K_{ab} 
\nonumber \\ 
& & + N [ \frac{1}{2} \gamma_5 \delta^c_k \delta^a_i \delta^b_j 
+ \frac{1}{2} \gamma_6 \delta^c_k g^{ab} g_{ij} \nonumber\\
& & + \frac{1}{2} \gamma_7 g^{ca} \delta^b_{(i} g_{j)k} 
+ \frac{1}{2} \gamma_8 g^{ab} \delta^c_{(i} g_{j)k} \nonumber \\
& & + \frac{1}{2} \gamma_9 g^{ca} \delta^b_k g_{ij} ] \partial_c g_{ab}.
\label{e:ppD}
\eeqa
Hyperbolicity is well established for the standard KST system, 
$\gamma_5 = \gamma_6 = ... = \gamma_9 = 0$, so as in 
Sec.~\ref{s:illustration}, it is best to seek a linear change of variables to 
reduce the constraint-damped system to the standard system.  Continuing the 
analogy with the scalar field, define
\beq
\bar K_{ij} := K_{ij} - \frac{1}{2} \gamma_5 g_{ij}.
\eeq
Then the equation for $\hat \partial_0 \bar K_{ij}$ has the same principal 
part as that for $\hat \partial_0 K_{ij}$.  The equation for $
\hat \partial_0 D_{kij}$ becomes
\beqa
{\hat \partial}_0 D_{kij} &\simeq& - N[ \delta^c_k \delta^a_i \delta^b_j 
- \frac{1}{2} \gamma_3 g^{ca} g_{k(i} \delta^b_{j)} \nonumber \\ 
& & - \frac{1}{2} \gamma_4 g^{ca} g_{ij} \delta^b_k + \frac{1}{2} \gamma_3 
g^{ab} g_{k(i} \delta^c_{j)} \nonumber \\ 
& & + \frac{1}{2} \gamma_4 g^{ab} g_{ij} \delta^c_k ] \partial_c \bar K_{ab} 
\nonumber \\ 
& & + N [ \frac{1}{2} (\gamma_6 - \frac{1}{2} \gamma_4 \gamma_5) 
\delta^c_k g^{ab} g_{ij} \nonumber\\
& & + \frac{1}{2} (\gamma_7 + \frac{1}{2} \gamma_3 \gamma_5) g^{ca} 
\delta^b_{(i} g_{j)k} \nonumber \\
& &+ \frac{1}{2} (\gamma_8 - \frac{1}{2} \gamma_3 \gamma_5) g^{ab} 
\delta^c_{(i} g_{j)k} \nonumber \\
& & + \frac{1}{2} (\gamma_9 + \frac{1}{2} \gamma_4 \gamma_5) g^{ca} 
\delta^b_k g_{ij} ] \partial_c g_{ab}.
\eeqa
If, for arbitrary $\gamma_5$, the further parameters are fixed as
\beqa
\gamma_6 &=& \frac{1}{2} \gamma_4 \gamma_5 \label{e:gm6fix}\\
\gamma_7 &=& - \frac{1}{2} \gamma_3 \gamma_5 \\
\gamma_8 &=& \frac{1}{2} \gamma_3 \gamma_5 \\
\gamma_9 &=& - \frac{1}{2} \gamma_4 \gamma_5 \label{e:gm9fix},
\eeqa
then the principal system, in the transformed variables, loses any reference 
to $\gamma_5, ... , \gamma_9$.
\beqa
{\hat \partial}_0 g_{ij} &\simeq& 0,\label{e:ppgtrans}\\
{\hat \partial}_0 \bar K_{ij} &\simeq& N[g^{ab} \delta^c_i \delta^d_j 
- (1+\gamma_2) g^{ad}\delta^b_{(i}\delta^c_{j)} \nonumber \\ 
& & - (1-\gamma_2) g^{bc}\delta^a_{(i}\delta^d_{j)} + (1+2{\gamma_0}) 
g^{cd}\delta^a_{(i}\delta^b_{j)} \nonumber \\ & & - \gamma_1 g^{ad} g^{bc} 
g_{ij} + \gamma_1 g^{ab} g^{cd} g_{ij} ] \partial_a D_{bcd} ,
\label{e:ppKtrans}\\
{\hat \partial}_0 D_{kij} &\simeq& - N[ \delta^c_k \delta^a_i \delta^b_j 
- \frac{1}{2} \gamma_3 g^{ca} g_{k(i} \delta^b_{j)} \nonumber \\ 
& & - \frac{1}{2} \gamma_4 g^{ca} g_{ij} \delta^b_k 
+ \frac{1}{2} \gamma_3 g^{ab} g_{k(i} \delta^c_{j)} \nonumber \\ 
& & + \frac{1}{2} \gamma_4 g^{ab} g_{ij} \delta^c_k ] \partial_c 
\bar K_{ab}.\label{e:ppDtrans}
\eeqa
This is the principal part of the standard KST system.  So for any value of 
the parameter $\gamma_5$, with $\gamma_6, ... , \gamma_9$ fixed 
by Eqs.~\eqref{e:gm6fix} -- \eqref{e:gm9fix}\footnote{This requirement can be 
weakened somewhat.  There is one further degree of freedom, shared between 
$\gamma_6$ and $\gamma_8$, which will still preserve hyperbolicity.  However 
when this degree of freedom is utilized, the simple argument used here must 
be replaced either by a somewhat more subtle argument, or a significantly 
more laborious one.  We have not yet found a use for this further degree of 
freedom, so here we restrict attention to the simpler case.}, 
the hyperbolicity of our modified 
system is the same as that of the corresponding standard KST system.

\section{Hyperbolicity of the constraint evolution}
\label{s:const_hyp}

Let us now turn our attention to the evolution of the constraint fields in our 
modified KST evolution system.  Given the definitions of the constraints in 
terms of the fundamental dynamical fields, an evolution system for the 
constraints follows from our fundamental evolution equations.  It is 
important, for the construction of constraint-preserving boundary conditions, 
that this system also be symmetric-hyperbolic ~\cite{Kidder2005, Stewart1998, 
Calabrese2003}.  The 
principal part of this evolution system can be expressed as,
\begin{eqnarray}
\hat \partial_0{\cal C} &\simeq & 
-\frac{1}{2}(2-\gamma_3+2\gamma_4) N g^{ij} \partial_i{\cal C}_j\nonumber\\
&&+\frac{1}{2}(\gamma_8-2\gamma_6-\gamma_5) N g^{ij} 
\partial_i{\cal C}^1_{j}\nonumber\\
&&+\frac{1}{2}(\gamma_7-2\gamma_9+\gamma_5) N g^{ij} 
\partial_i{\cal C}^2_{j},\label{e:hampp}\\
\hat \partial_0{\cal C}_i &\simeq & - (1+2\gamma_1) N 
\partial_i {\cal C}\nonumber\\
&&+\frac{1}{2} N g^{kl}g^{ab}\bigl[(1-\gamma_2)\partial_k{\cal C}_{labi}
+(1+\gamma_2)\partial_k{\cal C}_{ailb}\nonumber\\
&&\quad\qquad\qquad-(1+2{\gamma_0})\partial_k{\cal C}_{liab}\bigr],
\label{e:mompp}\\
\hat \partial_0{\cal C}_{kij} &\simeq & 0 ,\label{e:c3pp}\\
\hat \partial_0{\cal C}_{abij} &\simeq & - \frac{1}{2} N \gamma_3 
g_{i[a} \partial_{b]} {\cal C}_{j} - \frac{1}{2} N \gamma_3 
g_{j[a} \partial_{b]} {\cal C}_{i} \nonumber \\
&&- N \gamma_4 g_{ij} \partial_{[b} {\cal C}_{a]} - N \gamma_6 g_{ij} 
\partial_{[b} {\cal C}^1_{a]} \nonumber \\
&&- \frac{1}{2} N \gamma_7 g_{i[a} \partial_{b]} {\cal C}^2_{j} 
- \frac{1}{2} N \gamma_7 g_{j[a} \partial_{b]} {\cal C}^2_{i} \nonumber \\
&&- \frac{1}{2} N \gamma_8 g_{i[a} \partial_{b]} {\cal C}^1_{j} 
- \frac{1}{2} N \gamma_8 g_{j[a} \partial_{b]} {\cal C}^1_{i} 
\nonumber \\
&&- N \gamma_9 g_{ij} \partial_{[b} {\cal C}^2_{a]}, \label{e:c4pp}
\end{eqnarray}
where the four-index object ${\cal C}_{klij} := 2 \partial_{[k}D_{l]ij}$ 
is considered an independent constraint, 
so that the constraint evolution system is first-order.  

The inclusion of the terms proportional to $\gamma_5, ... , \gamma_9$ in 
Eqs.~\eqref{e:hampp} -- \eqref{e:c4pp} has seriously complicated 
this system.  Let us consider, however, the case considered above, 
where $\gamma_5$ is arbitrary, and the other parameters are fixed by 
Eqs.~\eqref{e:gm6fix} -- \eqref{e:gm9fix}.  In this case, a number of 
remarkable simplifications occur and the above system can be written as

\begin{eqnarray}
\hat \partial_0{\cal C} &\simeq & 
-\frac{1}{2}(2-\gamma_3+2\gamma_4) N g^{ij} \partial_i \bar{\cal C}_j,
\label{e:hampp2}\\
\hat \partial_0 \bar {\cal C}_i &\simeq & - (1+2\gamma_1) N 
\partial_i {\cal C}\nonumber\\
&&+\frac{1}{2} N g^{kl}g^{ab}\bigl[(1-\gamma_2)\partial_k{\cal C}_{labi}
+(1+\gamma_2)\partial_k{\cal C}_{ailb}\nonumber\\
&&\quad\qquad\qquad-(1+2{\gamma_0})\partial_k{\cal C}_{liab}\bigr],
\label{e:mompp2}\\
\hat \partial_0{\cal C}_{kij} &\simeq & 0 ,\label{e:c3pp2}\\
\hat \partial_0{\cal C}_{abij} &\simeq & - \frac{1}{2} N \gamma_3 
g_{i[a} \partial_{b]} \bar {\cal C}_{j} - \frac{1}{2} N \gamma_3 
g_{j[a} \partial_{b]} \bar {\cal C}_{i} \nonumber \\
&&- N \gamma_4 g_{ij} \partial_{[b} \bar {\cal C}_{a]}, \label{e:c4pp2}
\end{eqnarray}
defining the new combination $\bar {\cal C}_k := {\cal C}_k + \frac{1}{2} 
\gamma_5 {\cal C}^1_k - \frac{1}{2} \gamma_5 {\cal C}^2_k$.  

This constraint evolution system has the same principal part as the standard 
KST constraint system.  Thus, when the parameters are chosen by Eqs.
\eqref{e:gm6fix} -- \eqref{e:gm9fix}, the hyperbolicity of the fundamental 
and constraint evolution systems are independent of the parameter $\gamma_5$, 
so our modifications do not alter the hyperbolicity of these systems.

\section{Stability of constraint fields under free evolution}
\label{s:ModeAnalysis}

Analyzing the stability of the constraint evolution system in generic 
simulations is essentially no different than the full numerical relativity 
problem itself.  In order to get some handle, at the analytical level, on 
the effect of our modifications, we consider constraint 
violating perturbations of Minkowski spacetime.  Obviously these estimates 
will not be completely relevant in simulations of interest, but at least 
in the limit of short-wavelength perturbations, the dependence on the 
spacetime background should be minimal.  In this sense, stability of 
short-wavelength constraint-violating perturbations of Minkowski spacetime 
is a necessary condition for constraint damping in general.  And while our 
analysis of long-wavelength modes may not be directly relevant for evolutions 
of curved spacetime, unstable long-wavelength modes should at least be 
disconcerting, as a signal that instabilities are likely in general 
simulations.  

This analysis involves the full (not just principal) constraint 
evolution system, linearized about the limit that $g_{ij} = \delta_{ij}$, 
$K_{ij} = D_{kij} = 0$, $N = 1$, $N^i = 0$.  In this context, the full 
constraint evolution system becomes:
\beqa
\partial_t {\cal C} &=& - \frac{1}{2} (2-\gamma_3+2\gamma_4) \delta^{ij} 
[\partial_i {\cal C}_j \nonumber + \frac{1}{2} \gamma_5 \partial_i {\cal 
C}^1_j \\
&& \quad\qquad\qquad\qquad\qquad - \frac{1}{2} \gamma_5 
\partial_i {\cal C}^2_j],\label{e:Hamevfull}\\
\partial_t {\cal C}_i &=& - (1+2\gamma_1) \partial_i {\cal C}\nonumber\\
&&+\frac{1}{2} \delta^{kl}\delta^{ab} \bigl[(1-\gamma_2) 
\partial_k\partial_{[a}{\cal C}_{l]bi}
+(1+\gamma_2) \partial_k \partial_{[i} {\cal C}_{a]lb}\nonumber\\
&&\quad\qquad\qquad-(1+2{\gamma_0})\partial_k \partial_{[i}{\cal 
C}_{l]ab}\bigr] ,\label{e:Momevfull}\\
\partial_t {\cal C}_{kij} &=& - \gamma_5 {\cal C}_{kij} \nonumber \\
&& - \gamma_3 \delta_{k(i} {\cal C}_{j)} - \gamma_4 \delta_{ij} 
{\cal C}_k \nonumber \\
&& - \frac{1}{2} \gamma_3 \gamma_5 \delta_{k(i} {\cal C}^1_{j)} 
- \frac{1}{2} \gamma_4 
\gamma_5 \delta_{ij} {\cal C}^1_k \nonumber \\
&& + \frac{1}{2} \gamma_3 \gamma_5 \delta_{k(i} {\cal C}^2_{j)} 
+ \frac{1}{2} \gamma_4
\gamma_5 \delta_{ij} {\cal C}^2_k. \label{e:C3evfull}
\eeqa
Notice that we are no longer considering ${\cal C}_{klij}$ an independent 
constraint field.  In actual evolutions, where the fundamental fields are 
evolved, not the constraints, the three- and four-index constraints 
satisfy the identity
\beq
{\cal C}_{klij} = \partial_{[l} {\cal C}_{k]ij}.
\eeq
Violations of this identity will not appear in evolutions.

Now the above system is simplified by resolving all constraint fields into 
Fourier modes.  This has the formal effect of replacing all spatial 
derivatives $\partial_j$ with $- i k_j$, an imaginary unit times a 
propagation vector $k_j$.  The result is a system of coupled ODEs for the 
various 
constraint modes $c^A(k_i, t)$:
\beq
\partial_t c^A = M^A_B c^B.
\eeq
Each eigenvector of $M^A_B(k_i)$ evolves as $\exp(s t)$ for 
some $s(k_i)$.  The real part of $s$ is the rate of exponential growth (or 
damping, if negative) for the corresponding mode.  Due to the rotational 
invariance of the 
problem, these eigenvalues should depend only on the magnitude of $k_i$, so 
the propagation vector is decomposed as $k_i = k n_i$ where $n_i$ is a unit 
vector.

This eigenvalue problem naturally reduces into subspaces according to various 
possible spin weights about the propagation direction $n_i$.  There is a 
five-dimensional space of longitudinal modes: 
$\{ {\cal C}, {\cal C}_n, {\cal C}^1_n, {\cal C}^2_n, {\cal C}_{nnn} \}$, 
where ${\cal C}_n := n^i {\cal C}_i$, etc.  There is also a five-dimensional 
space of transverse vector modes: 
$\{ {\cal C}_I, {\cal C}^1_I, {\cal C}^2_I, {\cal C}_{Inn}, 
{\cal C}_{nnI} \}$, 
where capital latin indices now refer to a two-dimensional vector basis 
orthogonal to $n^i$.  The remaining constraint fields, with higher spin 
weight, are represented among the various projections of the totally tracefree 
part of ${\cal C}_{kij}$.  A glance at Eq.~\eqref{e:C3evfull} shows 
that all of these high spin-weight fields propagate trivially with 
$s = -\gamma_5$ independent of wavelength.  They are therefore damped 
exponentially on the timescale $\gamma_5^{-1}$ for positive $\gamma_5$.  The 
longitudinal and transverse constraint modes require more careful 
consideration.

\subsection{Transverse vector constraint modes}

The growth rates of the transverse vector modes are related to the 
eigenvalues of a five-by-five matrix.  Three of these eigenvalues simply 
equal $-\gamma_5$.  The remaining two are solutions of a quadratic 
equation, and depend on wavelength as:

\beq
s(k) = - \frac{1}{2} \gamma_5 \Gamma \pm \sqrt{\frac{1}{4} \gamma_5^2 \Gamma^2 
- v_2^2 k^2}, \label{e:TransverseDamp}
\eeq
where we define the convenient shorthand
\beq
\Gamma := \frac{1}{2} (2 - \gamma_3 + 2 \gamma_4),  \label{e:Gamma}
\eeq
and $v_2$ is one of the characteristic speeds of the KST system (relative to 
hypersurface-normal observers), 
\beq
v_2^2 := \frac{1}{8} \gamma_3 (1-3\gamma_2-4\gamma_0)-
\frac{1}{4}\gamma_4(1+6\gamma_0).  
\eeq
Notice that  one mode is undamped in the 
long-wavelength ($k \rightarrow 0$) limit, where one root in 
Eq.~\eqref{e:TransverseDamp} becomes zero.  This is not 
surprising: other constraint-damped 
representations of the Einstein system have the same 
property~\cite{Brodbeck1999, Gundlach2005, Lindblom2005}.  
In practice, long wavelength constraint 
modes should be killed off by proper constraint-preserving boundary 
conditions.  

In the short wavelength ($k \rightarrow \infty$) limit, 
the dispersion relation becomes
\beq
s(k) \rightarrow - \frac{1}{2} \gamma_5 \Gamma \pm i v_2 k .
\label{e:TransverseLimit}
\eeq
These represent propagating modes of the constraint system, damped at short 
wavelength on the timescale $(\frac{1}{2} \gamma_5 \Gamma)^{-1}$.  Notice the 
significance of the constant $\Gamma$.  Most of the modes require 
$\gamma_5 > 0$ for damping, so the damping of the modes referred to 
in Eq.~\eqref{e:TransverseLimit}
requires $\Gamma > 0$ as well.  Thus, the damping condition places a new 
restriction on the standard KST parameters $\{\gamma_0, ..., \gamma_4\}$, 
beyond the conditions they must satisfy for the system to be hyperbolic.

\subsection{Longitudinal constraint modes}

The longitudinal modes again involve the eigenvalues of a five-by-five 
matrix.  In this case 
two of the eigenvalues are simply $-\gamma_5$.  The rest are the roots of the 
cubic polynomial
\beq
s^3 + \gamma_5 \Gamma s^2 + k^2 v_3^2 s + k^2 \gamma_5 \Gamma (1 + 2 \gamma_1) 
= 0,\label{e:polynomial}
\eeq
where $v_3$ is another characteristic speed, given by
\beq
v_3^2 := \frac{1}{2}(1+2\gamma_1)(2-\gamma_3+2\gamma_4) 
- \frac{1}{2}\gamma_2\gamma_3. \label{e:v3}
\eeq
Rather than giving complicated analytic expressions for the roots of this 
polynomial, we simply consider asymptotic limits in $k$.  First, in the 
long-wavelength ($k=0$) limit, two roots vanish and the third is 
$ - \gamma_5 \Gamma$.  This is very similar to the long-wavelength behavior 
of the vector modes.

In the short-wavelength limit, the polynomial becomes singular.  The terms 
proportional to $k^2$ dominate the polynomial, leaving a linear 
equation.  The root of this linear equation, 
$s = - v_3^{-2} \gamma_5 \Gamma (1+2 \gamma_1)$, is the {\em regular root} 
of the polynomial in this limit.  The two remaining roots disappear
in the limit $k \rightarrow \infty$.  These singular roots 
correspond to traveling modes, with imaginary part linear in $k$ in this 
limit.  They can be found by substituting for $s$ a power series in $k$, 
$s = s_1 k + s_0 + s_{-1} k^{-1} + ...$ in the above polynomial and solving 
the resulting polynomial order-by-order in $k$ for the coefficients $s_i$.  
The result is
\begin{equation}
s(k) = - \frac{1}{2} \gamma_5 \Gamma \left(1 - 
\frac{1+2 \gamma_1}{v_3^2}\right) \pm i v_3 k + {\cal O}(k^{-1}).  
\label{e:slowestrate}
\end{equation}
So the damping of the traveling longitudinal modes requires that 
$v_3^2 > (1+2\gamma_1)$ when the transverse modes are damped as well.  

In summary, the damping of short-wavelength constraint-violating modes 
requires that the rates
\beqa
r_0 &:=& \gamma_5 \label{e:r0}\\
r_1 &:=& \frac{1}{2} \gamma_5 \Gamma \label{e:r1}\\
r_2 &:=& v_3^{-2} \gamma_5 \Gamma (1+2\gamma_1) \label{e:r2}\\
r_3 &:=& \frac{1}{2} v_3^{-2} \gamma_5 \Gamma (v_3^2 - 1 - 2 \gamma_1) 
\label{e:r3}
\eeqa
be positive, where $\Gamma$ is defined by Eq.~\eqref{e:Gamma} and 
$v_3$ by Eq.~\eqref{e:v3}.

\section{Choosing parameters}
\label{s:params}

Before proceeding with numerical tests, values must be fixed for the free 
parameters.  The parameters associated with the constraint damping terms are 
reasonably well set.  The overall damping timescale is set by $1/\gamma_5$, 
and this can be chosen to be any positive number.  The other new 
parameters are determined by Eqs.~\eqref{e:gm6fix} -- \eqref{e:gm9fix}.  The 
original 
KST parameters should be chosen in accord with hyperbolicity conditions for 
the fundamental and constraint evolution systems, as well as the conditions 
that the damping rates of Eqs.~\eqref{e:r0} -- \eqref{e:r3} be positive.

The hyperbolicity conditions are quite complicated when considered in full 
generality.  To make the situation more tractable, here we restrict 
attention to the subset of parameter space in which all characteristic speeds 
are equal to zero or unity, relative to hypersurface-normal observers.  The 
hyperbolicity conditions in this subset of the parameter space are spelled 
out in Appendix B of Ref.~\cite{Kidder2005}, following work 
in Ref.~\cite{Lindblom2003}.  The parameters $\gamma_0$, $\gamma_3$ 
and $\gamma_4$ 
are fixed in terms of $\gamma_1$ and $\gamma_2$ by the the conditions on the 
characteristic speeds:
\beqa
\gamma_0 &=& \frac{1}{2}\\
\gamma_3 &=& \frac{-8}{4\gamma_2 + (5 + 3 \gamma_2)(1+2\gamma_1)}\\
\gamma_4 &=& \frac{1 - \gamma_2 - (1+2\gamma_1)(5+3\gamma_2)}{4\gamma_2 
+ (5 + 3 \gamma_2)(1+2\gamma_1)}.
\eeqa
The fundamental evolution system is then symmetric-hyperbolic so long as the 
following inequalities are satisfied:
\beqa
&-\frac{5}{3} < \gamma_2 < 0 \\
&4 \gamma_2 + (1+2 \gamma_1)(5+3\gamma_2) \neq 0.
\eeqa

Constraint damping requires that the rates $r_i$ of 
Eqs.~\eqref{e:r0} -- \eqref{e:r3} be 
positive.  This in turn requires that $\Gamma>0$ and that 
\beq
0 < 1+2\gamma_1 < v_3^2.
\eeq
For the numerical simulations presented in the next section, 
$\gamma_1 = -1/4$.  The parameter 
$\Gamma := (1/2)(2 - \gamma_3 + 2 \gamma_4)$ can be expressed in terms of 
$\gamma_2$ using the above expressions for $\gamma_3$ and $\gamma_4$:
\beq
\Gamma = \frac{5 + 3 \gamma_2}{4 \gamma_2 + (5+3\gamma_2)(1+2\gamma_1)} = 
\frac{10 + 6 \gamma_2}{5 + 11 \gamma_2},
\eeq
where the last equality is restricted to the case 
$\gamma_1 = -1/4$.  In the allowable region for $\gamma_2$, $\Gamma$ can be 
set equal to any value greater than 2.  Here we choose $\Gamma = 5/2$.

The various parameters, and the associated growth rates, come out to:
\beqa
\gamma_0 &=& \frac{1}{2}, \label{e:gamma_0}\\
\gamma_1 &=& -\frac{1}{4}, \label{e:gamma_1}\\
\gamma_2 &=& -\frac{5}{43}, \label{e:gamma_2}\\
\gamma_3 &=& -\frac{43}{10}, \label{e:gamma_3}\\
\gamma_4 &=& -\frac{52}{80}, \label{e:gamma_4}\\
r_0 &=& \gamma_5,\\
r_1 &=& \frac{5}{4} \gamma_5,\\
r_2 &=& \frac{5}{4} \gamma_5,\\
r_3 &=& \frac{5}{8} \gamma_5.
\eeqa

These parameters satisfy all the necessary conditions for constraint damping 
in perturbations of flat spacetime, as well as those for symmetric-hyperbolic 
propagation of the fundamental evolution fields.  Unfortunately, these 
parameters do not satisfy all of the necessary conditions for 
symmetric-hyperbolic constraint propagation.  In Ref.~\cite{Kidder2005} it was 
shown that when the adjustable 
characteristic speeds are all set to unity, the symmetric hyperbolicity 
conditions on the fundamental and constraint evolution systems collude to 
require that $1 + 2 \gamma_1 < 0$, a direct conflict with our damping 
conditions.  Unfortunately, this conflict does not appear to be an artefact 
of our condition that all adjustable characteristic speeds are equal to one.  
Monte Carlo searches over the entire available parameter space have not 
provided us with any examples of systems with constraint damping along with 
symmetric-hyperbolic propagation of the fundamental and constraint fields.

In principle, this conflict is very serious.  At timelike boundaries of the 
simulation domain, conditions must be imposed on fields entering the 
computational grid.  These boundary conditions should be compatible with the 
constraint 
equations.  In Ref.~\cite{Kidder2005}, such boundary conditions were 
presented.  These conditions control the 
growth of a certain norm of the constraint fields.  In the case of the 
parameters used here, this norm is not positive-definite, so control of the 
norm does not necessarily imply control of the constraint fields themselves.

In practice, the damage done by this conflict can only be assessed with 
numerical simulations.  While the constraint evolution is not 
symmetric-hyperbolic, it is strongly hyperbolic, so the boundary conditions 
of Ref.~\cite{Kidder2005} can still be applied, even if they may not have 
all of the desired effects.  In 
fact, the numerical results of the following section demonstrate that 
constraint-preserving boundary conditions are quite effective in these 
simulations.  Perhaps this can be explained heuristically by the fact that 
the ``timelike'' 
degree of freedom in the constraint evolution (the one whose violations could 
compensate, in the indefinite norm, for violations of the other constraints) 
is very well controlled by the constraint damping.  

It should also be noted that without the constraint damping terms, the 
particular parameter set used here leads to very unstable evolutions.  In the 
following section, we will not make comparisons with the undamped 
case, $\gamma_5 = 0$, as those cases immediately become unstable.  This 
could be due, in part, to the lack of symmetric-hyperbolic constraint 
evolution.  At any rate, when the constraint damping terms are included, the 
evolutions become remarkably stable.

\section{Numerical Tests}
\label{s:num_tests}

The following numerical tests were carried out using the Spectral Einstein 
Code developed over the last few years by the numerical relativity groups at 
Cornell and Caltech.  The code uses multidomain 
pseudospectral collocation methods to resolve the fields in space with 
exponential accuracy.  Integration in time is implemented by the method of 
lines, using in this case a fourth-order Runge-Kutta scheme.  More details on 
this code and its remarkable accuracy can be found in  Ref.~\cite{Boyle2006} 
and references therein.

The spectral representation of the computed fields is done in accordance with 
the topology of the spatial domain.  The present simulations are 
of a single 
Schwarzschild black hole, in 
Kerr-Schild~\cite{Chandra83} coordinates.  
The spatial 
domain is made up of a family of concentric thick spherical shells.  The 
fields are therefore resolved into spherical harmonics in the angular 
directions, multiplied by Chebyshev polynomials in the radial direction.  
The 
innermost boundary is inside the black hole horizon, so no boundary condition 
is needed there.  At the outermost boundary, the constraint-preserving 
boundary conditions presented in Ref.~\cite{Kidder2005} are used.  
As in Ref.~\cite{Kidder2005}, tensor 
spherical harmonic components of the four highest $l$ values are discarded 
after each time step.  No filtering appears to be necessary in the radial 
direction.

\begin{figure} 
\begin{center}
\includegraphics[width=3in]{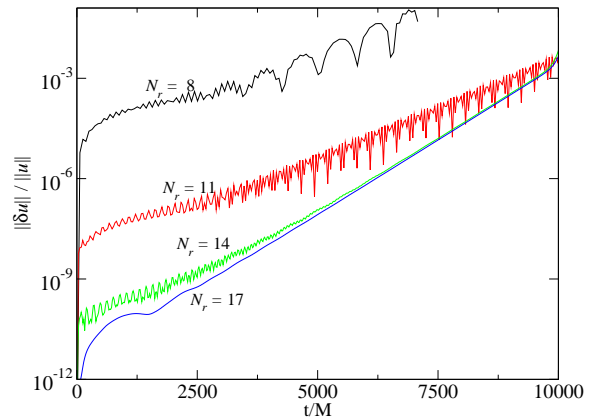}
\end{center}
\caption{Norm of the error $||\delta u||/||u||$, relative to the reference 
solution, on a fixed domain extending from minimum coordinate 
radius $1.9 M$ to maximum $41.9 M$.  The domain is broken into eight shells 
each of thickness $5 M$ and radial resolution $N_r$, chosen on four different 
runs as $N_r = 8, 11, 14, 17$.  The 
constraint damping terms presented in all of these simulations have 
$\gamma_5 = 0.6 M^{-1}$.
\label{ErrorConvergence}}
\end{figure}

\begin{figure} 
\begin{center}
\includegraphics[width=3in]{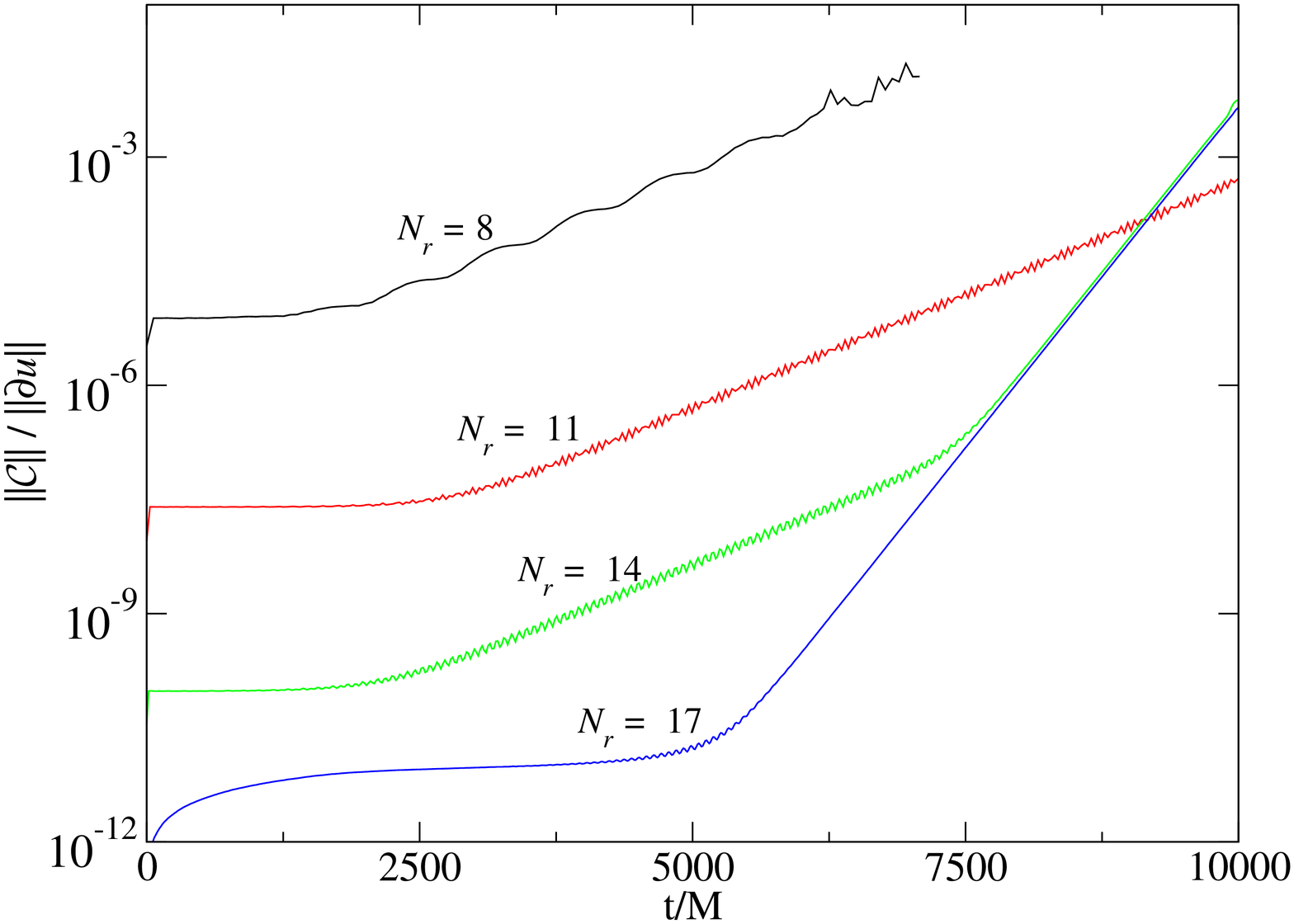}
\end{center}
\caption{Constraint norm $||{\cal C}||/||\partial u||$, for the same runs 
plotted in Fig.~\ref{ErrorConvergence}.  
\label{ConstraintConvergence}}
\end{figure}

Figures~\ref{ErrorConvergence} and~\ref{ConstraintConvergence} demonstrate 
the stability and exponential convergence of these simulations.  
Figure~\ref{ErrorConvergence} is a plot of the error norm: 
\beq
||\delta u||^2 := \int \left( \delta g^{ij} \delta g_{ij} + \delta K^{ij} 
\delta K_{ij} + \delta D^{kij} \delta D_{kij} \right) dV, 
\eeq
measuring the difference between the computed solution and the reference 
Kerr-Schild geometry.  Figure~\ref{ConstraintConvergence} shows a 
positive-definite norm of the constraint fields:
\beq
||{\cal C}||^2 := \int \left( {\cal C}^2 + \frac{1}{3}{\cal C}^i{\cal C}_i + 
\frac{1}{18}{\cal C}^{kij}{\cal C}_{kij} + \frac{1}{18}{\cal C}^{klij}
{\cal C}_{klij} \right) dV.
\eeq
We normalize these 
quantities, dividing by norms that involve similar terms, but that should not 
be expected to vanish.  The error norm is divided by the overall solution 
norm: 
\beq
||u||^2 := \int \left( g^{ij} g_{ij} + K^{ij} K_{ij} 
+ D^{kij} D_{kij} \right) dV,
\eeq
and the constraint norm is divided by a similar norm of the first derivatives 
of the computed fields:
\beqa
||\partial u||^2 &:=& \int ( g^{kc} g^{ia} g^{jb} \partial_k g_{ij} 
\partial_c g_{ab}  \nonumber \\
& & + g^{kc} g^{ia} g^{jb} \partial_k K_{ij} \partial_c K_{ab} \nonumber \\
& & + g^{ld} g^{kc} g^{ia} g^{jb} \partial_l D_{kij} \partial_d D_{cab} ) dV.
\eeqa
All indices are raised and lowered with the computed metric $g_{ij}$.

Here, the inner 
(excision) boundary is at $1.9 M$, and the outer boundary is at $41.9 M$.  
This domain is divided into eight subdomains, each of coordinate thickness 
$5 M$.  This is the same domain used in Ref.~\cite{Kidder2005}.  
Note that the convergence stops at the highest resolution presented here, 
overtaken by exponential growth that is not yet apparent in the constraint 
fields shown in Fig.~\ref{ConstraintConvergence}.  In Ref.~\cite{Kidder2005}, 
a ``gauge instability'' was mentioned, 
associated with one particular boundary condition.  Presumably, this is the 
same instability apparent in Fig.~\ref{ErrorConvergence}, in which case it 
could be 
expected that convergence would improve as the location of the outer boundary 
is moved farther into the asymptotic regime.

\begin{figure} 
\begin{center}
\includegraphics[width=3in]{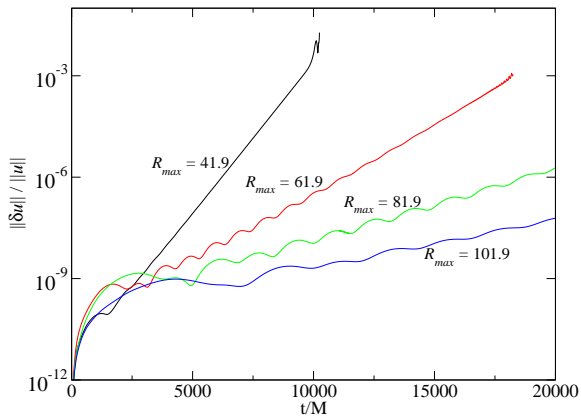}
\end{center}
\caption{Error norm $||\delta u||/||u||$ for runs in the 
constraint-damped system with 
outer boundary at $r =$ 41.9 $M$, 61.9 $M$, 81.9 $M$, 101.9 $M$.  
The long-term growth of the 
error norm occurs exponentially on a timescale proportional to the 
square of the coordinate position of the outer boundary.  
\label{Boundary_Error}}
\end{figure}

\begin{figure} 
\begin{center}
\includegraphics[width=3in]{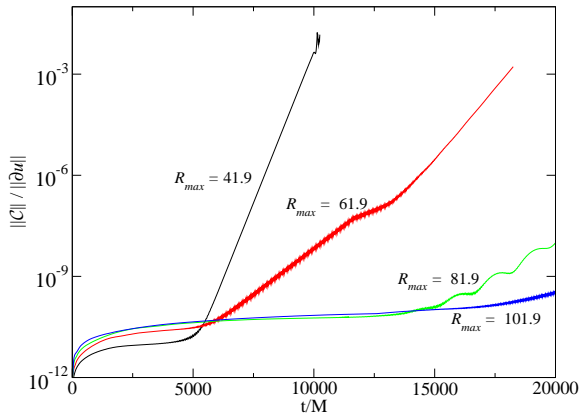}
\end{center}
\caption{Constraint norm $||{\cal C}||/||\partial u||$ of the same 
runs as those in Fig.~\ref{Boundary_Error}.  The constraints grow 
as $t^{1/2}$ until eventually 
driven exponentially by the overall loss of accuracy demonstrated in 
Fig.~\ref{Boundary_Error}.  
\label{Boundary_Constr}}
\end{figure}

As a test of this hypothesis, the highest-resolution run in 
Fig.~\ref{ErrorConvergence} was repeated on 
larger domains, keeping resolution fixed but adding extra subdomains to place 
the outer boundary at coordinate radii 61.9 $M$, 81.9 $M$, 101.9 $M$.  
Figure~\ref{Boundary_Error} demonstrates the improvement in the overall error 
norm.  Least-squares fitting of the data in that plot show that the 
late-term growth in this error occurs exponentially on a 
timescale proportional to the square of the coordinate position of the outer 
boundary.  Figure~\ref{Boundary_Constr} shows the growth of 
constraint energy in these simulations.  Until an exponential instability sets 
in, apparently triggered by the overall loss of accuracy of the simulation, 
the constraint fields grow roughly as the square root of coordinate time.  
On the largest domain, this slow growth persists beyond 15,000 $M$, when 
exponential growth takes over at a rate that would allow the simulation to 
survive until nearly 50,000 $M$.

\begin{figure}[t]
\begin{center}
\vspace{2mm}
\includegraphics[width=3in]{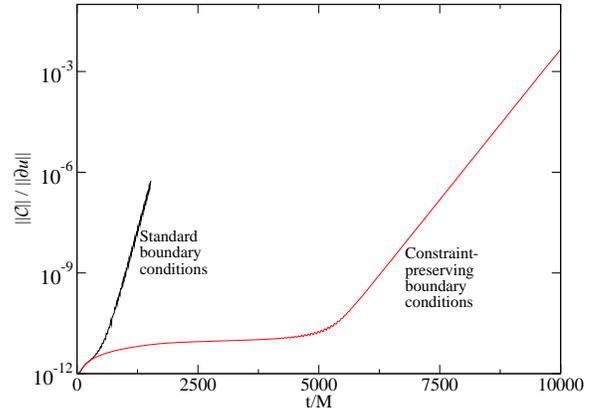}
\end{center}
\caption{Constraint norm $||{\cal C}||/||\partial u||$ of a black hole 
simulation with $R_{max} = 41.9$, $N_r = 17$ (in each of eight subdomains), 
and two different boundary conditions.  
\label{BC_Comparison}}
\end{figure}

It is of some interest to verify the effectiveness of the 
constraint-preserving boundary conditions used in these simulations.  As noted 
in the previous section, since the characteristic matrices of the constraint 
evolution system are symmetric only with respect to a Lorentzian norm, there 
is no reason to expect these conditions to control the influx of constraint 
violations.  In Fig.~\ref{BC_Comparison}, the simulation with 
$R_{max} = 41.9$ and $N_r = 17$ (in each subdomain) is repeated using 
conventional boundary conditions.  These boundary conditions freeze the 
incoming characteristic fields of the fundamental evolution system to their 
initial values.  These ``freezing'' boundary conditions control a 
positive-definite norm of the fundamental evolution fields, so the initial 
boundary value problem is known to be well-posed by standard theorems.  
However, the figure clearly demonstrates the superiority of the 
constraint-preserving boundary conditions in this context, not only for 
constraint satisfaction, but for overall stability.  Perhaps the effectiveness 
of the constraint preserving conditions is not robust, perhaps it will fail 
when the conditions are applied in more dynamical spacetimes.  This 
possibility is an important avenue for further investigation -- if this 
stability is found not to be robust, then either the spatial domain will need 
to be compactified to remove timelike boundaries, or further modification of 
the KST system will be needed to combine the constraint damping effects 
outlined here with truly symmetric hyperbolic constraint propagation.\\

\section{Discussion}

A generalization of the five-parameter KST 
systems was introduced, for use in numerical relativity.  The added 
parameter $\gamma_5$ 
supplies a timescale on which exponential damping can occur (or growth, if 
parameters are not chosen carefully).  
The hyperbolicity of the fundamental and constraint evolution systems is 
not changed by this modification, but the effect that the constraint damping 
has on perturbations of flat spacetime is partly dependent on the same 
parameters that determine hyperbolicity.  Parameters can be chosen such that 
all constraint modes are stable in perturbations of flat spacetime, but 
not when the constraint fields are required to evolve in a 
symmetric-hyperbolic manner.  Nevertheless, single black hole simulations 
using constraint-preserving boundary conditions are convergent, 
and what instabilities exist appear to be dominated by constraint-satisfying 
modes, associated with a gauge instability in the outer boundary condition.

\acknowledgments

I thank Lee Lindblom for extensive discussions and advice, and also Harald 
Pfeiffer and Mark Scheel, for help on the manuscript and for assistance on 
the use of their code.  Michael Boyle provided access to some input files and 
numerical results of Ref.~\cite{Boyle2006}, which were of great help in 
debugging and understanding the numerical issues.
The numerical calculations presented in this paper were performed with the 
Caltech/Cornell Spectral Einstein Code (SpEC) written primarily by 
Lawrence E. Kidder, Harald P. Pfeiffer and Mark A. Scheel.  This work was 
suppored in part by NSF grants PHY-0099568, PHY-0244906, PHY-0601459 
and DMS-0553302, NASA grants NAG5-12834 and NNG05GG52G, and a grant from the 
Sherman Fairchild Foundation.  Some of the numerical calculations leading to 
this paper were performed with the Tungsten cluster at NCSA.


\end{document}